\documentclass[journal=jpccck,manuscript=article]{achemso}
\usepackage{tikz,pgfplots}
\usepackage{amsmath}
\usepackage{subcaption}
\usepackage[T1]{fontenc}

\setkeys{acs}{articletitle = true}

\author{Frederike Jaeger}
\author{Omar K. Matar}
\author{Erich A. M\"{u}ller}
\email{e.muller@imperial.ac.uk}
\affiliation[Imperial Chem Eng]
{Department of Chemical Engineering, Imperial College London, London, UK}

\title{Transport Properties of Water Confined in a Graphene Nanochannel}


\begin{document}

\begin{abstract}
Equilibrium molecular dynamics simulations are used to investigate the effect of phase transitions on the transport properties of highly-confined water between parallel graphene sheets. An abrupt reduction by several orders of magnitude in the mobility of water is observed in strong confinement, as indicated by reduced diffusivity and increased shear viscosity values. The bulk viscosity, which is related to the resistance to expansion and compression of a substance, is also calculated, showing an enhancement compared to the bulk value for all levels of confinement. An investigation into the phase behaviour of confined water reveals a transition from a liquid monolayer to a rhombic `frozen' monolayer at nanochannel heights between 6.8-7.8 \AA; for larger separations, multilayer liquid water is recovered. It's shown how this phase transition is at the root of the impeded transport. 
\end{abstract}

\newpage
\section{Introduction}
Recently, the concept of being able to engineer materials for the fine control of water flow through nanochannels has gathered a lot of attention. Its application can be envisioned, for example, in membrane separation technology, where selectivity is achieved through both the size of the nanoscale pore or channel \cite{Corry2008,Cohen-Tanugi2012,Wang2012a,Kou2014,An2017}, or alternatively by manipulating the surface chemistry effectively mimicking mechanisms found in biological membranes  \cite{Corry2011,He2013,Konatham2013b,Chan2014}. Water, a substance which already displays anomalous properties in the bulk \cite{Ball2008,Stokely2008,Nilsson2015}, shows even more curious behaviour under confinement. Interesting structural and transport phenomena, such as single file diffusion \cite{Hummer2001,Kosztin2006,Bocquet2010,Wang2011,Gravelle2014} and ultra-fast flow through carbon nanotubes (CNTs) \cite{Majumder2005,Holt2006}, are only some examples of this.  The behaviour of water under confinement has been studied at length with a view of quantifying the extent of these phenomena in terms of the properties of the porous media such as geometry, surface interaction and in terms of the fluid's thermodynamic state \cite{Majumder2005,Holt2006,Thomas2009a,Su2012,Suk2013,Calabro2013,Ritos2014,Mattia2015,Kannam2017,Wang2017}.\\
The confinement of fluids between the walls of molecular-size pores induces important changes in the thermodynamic stability of the possible phases \cite{Gelb1999}. One of the striking structural effects observed for strongly confined water is the formation of ice-like structures at temperatures above the freezing point. High temperature freezing of water in carbon nanotubes (CNTs) has not only been observed in simulation studies \cite{Koga2001,Koga2002,Takaiwa2008,Pugliese2017,Raju2018}, but also experimentally \cite{Maniwa2005,Agrawal2017}. Agrawal et al. \cite{Agrawal2017} confirmed solidification of water inside $10.5$ \AA~diameter CNTs at temperatures above the boiling point (up to $411$ K). Here, freezing is highly sensitive to tube diameter, commensurate with simulation studies. The formation of ice phases has also been observed in water confined between two parallel walls, where, in addition to a dependence on the channel height \cite{Zangi2003,Zangi2003a,Zhao2014,Zhu2015}, ice phases are also strongly influenced by high lateral pressures \cite{Zhao2014,Zhu2015,Algara-Siller2015,Yang2017}. Whilst six different monolayer ice phases have been shown to emerge in confinement \cite{Zhao2014} (usually at separations of 6-7 \AA \cite{Mosaddeghi2012,Zhao2014,Algara-Siller2015,Yang2017}), at room temperature predominantly square-like or rhombic phases are reported \cite{Mosaddeghi2012,Zhao2014,Zhou2015,SobrinoFernandezMario2015,Yang2017}, supported experimentally by Algara-Siller et al. \cite{Algara-Siller2015}. The observation of ice phases is typically accompanied with a sharp drop in water mobility, both in CNTs \cite{Zheng2012,DaSilva2014,Mashl2003} and parallel 2d graphene nanochannels \cite{Hirunsit2007,Mosaddeghi2012,Neek-Amal2016}. This manifests itself through decreased diffusivities or in turn increased viscosities, often reaching values associated with low temperature \cite{Hirunsit2007} or high pressure \cite{Neek-Amal2016} water. \\
The unfavorable environment experienced by water under confinement in hydrophobic spaces, along with the strong directional fluid-fluid interactions, conspire to give these systems unique thermodynamic properties and are responsible for unexpected transport properties, not only in the frozen but also liquid regime. Well-cited experimental studies of water flow through CNTs reported large flow enhancements \cite{Majumder2005,Holt2006}, pointing towards higher mobility within the tube. Commensurate with this, computational studies of water confined in CNTs often report decreased viscosities, either calculated by applying a two-viscosity hydrodynamic model to the data \cite{Myers2011,Thomas2008,Thomas2010,Mattia2012} or by computing the viscosity directly \cite{Chen2008,Babu2011,Ye2011,Zhang2011,Kohler2016}. In some cases, the reported reduction in the viscosity is of orders of magnitude. However, for graphene or graphite nanochannels, often the opposite is observed. Mobility is sometimes decreased for strong confinement \cite{Hirunsit2007,Mosaddeghi2012,Liu2016b,Neek-Amal2016,Fang2018}, where only at channel heights of around four molecular diameters approximate bulk behaviour is recovered \cite{Hirunsit2007,Mosaddeghi2012}. Therefore, it is clear that the fluid-wall interaction alone does not govern the transport of fluids in confinement.\\
Whilst transport properties of molecular fluids are usually quantified through the diffusivity and the shear viscosity, another transport property, the bulk viscosity, characterizes fluid behavior. The bulk viscosity is typically neglected, even though it can take on values three orders of magnitude larger than the shear viscosity, as is e.g. the case for CO$_2$ \cite{Tisza1942,Emanuel1990,Rah2001,Cramer2012,Li2017,Jaeger2018}. The bulk viscosity describes how a fluid's compression and dilation affects transport. In the case of gases, it is entirely related to the relaxation of vibrational and rotational degrees of freedom \cite{Tisza1942,Monchick1963,Prangsma1973,Emanuel1990,Mayorga1997,Cramer2012,Li2017,Lin2017}, whilst for dense fluids it contains additional contributions due to long-range molecular interactions \cite{Gray1964,Hanley1976,Rah2001}. For some liquids, including water \cite{Jaeger2018}, the vibrational and rotational contributions are negligible, enabling the bulk viscosity to be obtained from fluctuations in the pressure tensor using molecular dynamics simulations. The effects of the bulk viscosity on fluid transport are not well explored, though it has been shown to influence the propagation of shock waves \cite{Weyl1949,Gilbarg1953,Emanuel1998, Mani2009}, heat transfer in hypersonic flows \cite{Emanuel1990,Emanuel1992}, and the vorticity in supersonic combustion \cite{Billet2008,Giovangigli2015}. For a detailed review and current computational methods and results, the reader is referred to our recent study \cite{Jaeger2018}. Computational studies of simple fluids in confinement have shown that large and spatially varying tangential pressures occur within the channel, much larger than system pressures in the corresponding bulk phase, leading to unusual phenomena such as freezing and pressure-driven chemical reactions \cite{Long2011,Coasne2014,Srivastava2017,Srivastava2018}. As the bulk viscosity is strongly related to pressure fluctuations, the effects of confinement on the pressure has the potential to impact the bulk viscosity significantly.\\
For a successful integration of graphene channels into nanoscale membrane applications \cite{Boretti2018,Seo2018,Tabish2018}, the sensitivity of water to the geometry and thermodynamic properties of the nanochannel needs to be explored. In this paper, the structure and dynamics of room temperature water under confinement between two graphene sheets is investigated using equilibrium molecular dynamics studies. A particular focus is placed on the bulk viscosity under confinement, which has only been studied for model fluids in the literature \cite{Goyal2013}. In order to isolate the contributions of the degree of confinement, water is kept at a constant average channel density regardless of channel height, allowing for a detailed analysis of the differences between bulk water at ambient pressure and temperature and its confined counter part. 

\section{Methods}
\subsection{Simulation setup}
\label{sec:conf:sim}

\begin{figure}
	\centering
	\includegraphics[width=\textwidth]{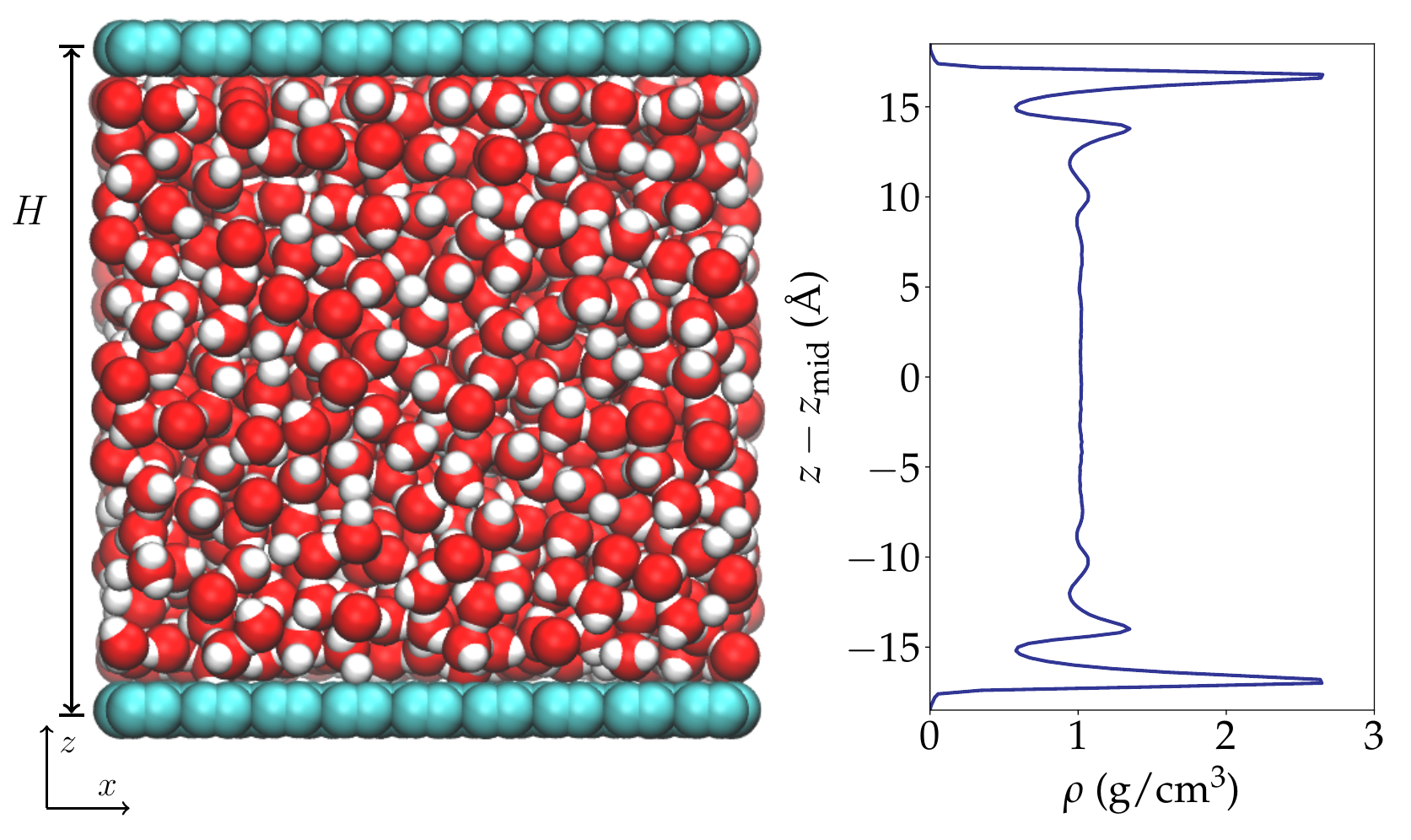}
	\caption{Simulation setup and density profile of water confined within a graphene nanochannel of height $H = 40$ \AA. The simulation temperature is $298$ K and the average channel density $1$ g/cm$^3$. The height is defined as the center-to-center distance of the carbon atoms.}
	\label{fig:cf:rhoreg}
\end{figure}

Water confined between two parallel graphene sheets is studied with molecular dynamics simulations using the Large-scale Atomic/Molecular Massively Parallel Simulator (LAMMPS) \cite{Plimpton1995}. The size of the simulation box is $(L_x, L_y, L_z) = (37.9, 32.8, H+5)$ \AA, where $H$ is defined as the distance between the centers of the carbon atoms in opposing graphene sheets. Periodic boundary conditions are applied in the $x$ and $y$ direction, whilst the $z$ direction remains fixed to avoid interactions of water molecules across boundaries. $H$ is varied between $6$ and $40$ \AA~in order to explore different states of confinement, from monolayer water to large separations where a significant bulk region is recovered at the center of the channel (see Fig. \ref{fig:cf:rhoreg}). For small separations, more finely spaced distances were examined, in order to observe the layering transition behavior of water. In order to calculate the fluid volume, an effective height of the channel, $H_{\mathrm{eff}}$, is defined as $H_{\mathrm{eff}} = H - \sigma_{\mathrm{CO}}$, where $\sigma_{\mathrm{CO}} = 3.19$ \AA~is the Lennard-Jones characteristic distance for the carbon-oxygen interaction \cite{Werder2003}. Consequently, the volume occupied by the fluid is given by $V_f = L_x \times L_y \times H_{\mathrm{eff}}$. This definition of $H_{\mathrm{eff}}$ has been used in the literature \cite{Mosaddeghi2012} to best match the average density in the channel to the density of the bulk regions of water for large separations. The channel average density in all cases is $\rho \simeq 1$ g/cm$^3$ in accordance with the bulk density of water. Additional simulations were performed at higher average channel densities. For each separation, three different starting configurations in terms of velocity seeds were probed.\\

Water is simulated using the SPC/E force field \cite{Berendsen1987}. SPC/E is chosen over more accurate but more computationally expensive models, such as TIP$4$P/2005 and TIP$4$P/ICE, in order to balance accuracy in the determination of transport properties with the associated computational cost. The water-carbon interaction used in this work is parametrized using the SPC/E model \cite{Werder2003}, a potential frequently used in water flow simulations \cite{Khademi2011,Zhu2013,Wei2014,Sam2017}. Previous studies have shown ice formation of monolayer SPC/E and TIP$4$P/2005 models with the same lattice constant \cite{Algara-Siller2015}, confirming the independence of these results with respect to the details of the force fields. The bonds and angles are constrained using the SHAKE algorithm \cite{Ryckaert1977}. The carbon-water interaction parameters by Werder et al. \cite{Werder2003}, optimized to reproduce the experimentally observed contact angle, are used, where $\epsilon_{\mathrm{CO}} = 0.00406$ eV and $\sigma_{\mathrm{CO}} = 3.19$ \AA. The chosen cut-off radius is $r_c = 13$ \AA, with the PPPM method \cite{Rajagopal1994} employed to evaluate long-range electrostatic interactions. The carbon atoms are uncharged and non-interacting. The graphene sheets consist of $486$ carbon atoms each, whilst the water phase comprises $122$ to $1536$ molecules, depending on the channel separation, $H$. In order to maintain a system temperature of $298$ K, an equilibration run of $3$ ns with a Nos\'{e}-Hoover thermostat \cite{Nose1984,Hoover1985} applied to both the graphene sheets and the water molecules with a damping constant of 0.05 ps is performed. The thermostat is subsequently removed and temperature control was ensured by applying a Berendsen thermostat \cite{Berendsen1984} with a damping term of 100 ps to the graphene sheets only, allowing for the water dynamics to be observed unimpeded. The carbon atoms are tethered to their relative position with a spring constant of $K_{\mathrm{tether}} =  4.3363$ eV/\AA$^2$ to hold the sheet in place and allow for vibrations to dissipate energy. After a further equilibration period of $1$ ns in the microcanonical ensemble, a production run is performed for $4$ ns.  \\

\subsection{Green-Kubo methods for transport calculations}

The velocity autocorrelation function (VACF) quantifies how strongly atom velocities at a certain time influence the velocities at later points. The normalized VACF is given by
\begin{equation}
C_{u,u}(t) =  \frac{1}{N_{\text{atoms}}}\sum_{i=1}^{N_{\text{atoms}}} \frac{\langle \mathbf{u}_i(0)\cdot \mathbf{u}_i(t) \rangle_m}{\langle \mathbf{u}_i(0)\cdot \mathbf{u}_i(0) \rangle_m},
\label{eq:tech:vacf}
\end{equation}
where all $N_{\text{atoms}}$ fluid atoms are considered in the averaging. $\langle\rangle_m$ is the ensemble average of the correlation for different time origins. Here, the velocity vector in the two dimensions parallel to the graphene sheets is given by $\mathbf{u}_{i} (t) = (u_x(t), u_y(t))$. The diffusion coefficient can be calculated using the mean-squared displacement (MSD), using
\begin{equation}
D_s = \frac{1}{2d} \lim_{t\rightarrow\infty}\frac{\langle\left[\mathbf{r}_{i}(t)-\mathbf{r}_{i}(0)\right]^2\rangle}{t},
\end{equation}
where the dimensionality is taken as $d=2$ so as to only take into account the MSD in the plane parallel to the graphene sheet. The slope of the MSD hence determines the diffusivity. Here, $\mathbf{r}_i$ is the position vector corresponding to the velocity vector given above, i.e. $\mathbf{r}_{i} (t) = (x(t), y(t))$.\\

The shear viscosity, $\eta$, is related to the time autocorrelation function of the off-diagonal elements of the stress tensor, $\mathcal{P}_{\alpha\beta}$, with $\alpha \neq \beta$, and is as such a tensor. In a homogeneous system all elements of this tensor, $\eta_{\alpha\beta}$, are equivalent and can be averaged in order to achieve more accurate results. However, in confinement this is not the case and the elements need to be carefully chosen for their validity. In this definition, $\alpha$ is the direction of the velocity in shear flow, and $\beta$ the direction of displacement. Previous work \cite{Liu2005} has shown that calculating $\eta_{\alpha z}$, where $z$ is the direction of confinement, produces unphysically low values. As such, only values of $\beta$ which are not the direction of the confinement are employed. We therefore calculate $\eta$ through
\begin{equation}
\eta \equiv \eta_{xy} = \frac{V_f}{k_BT}\sum_{\alpha\beta}\int_0^t\langle \mathcal{P}_{xy}(t^{\prime})\mathcal{P}_{xy}(0)\rangle dt^{\prime},
\end{equation}
where $V_f$ is the fluid volume as defined before.\\

Additionally, the bulk viscosity, $\kappa$, of confined water is calculated. In an isotropic system with short vibrational relaxation times, as is the case for water, $\kappa$ is calculated using the time autocorrelation function of the pressure. The question arises on how to apply this to a highly confined system. In the only known calculation of the bulk viscosity under confinement in the literature \cite{Goyal2013}, the correlation function is not calculated directly but expressed as a phenomenological analytic equation as a function of relaxation times accounting for structural and dynamical contributions. Confinement is accounted for by modifying the dynamical relaxation time to include a dependence on the distance from the confining wall, which is integrated over. A solution more closely related to the original definition in terms of correlation functions is to consider confined water as a $2d$ fluid, implying no expansion or dilation takes place in the $z$ direction. This allows us to define the instantaneous pressure as $\mathcal{P}(t)=(\mathcal{P}_{xx}+\mathcal{P}_{yy})/2$ and calculate its correlation over time, such that $\kappa$ is given by
\begin{equation}
\kappa = \frac{V_f}{k_BT}\int_0^t\langle \delta \mathcal{P}(t^{\prime})\delta \mathcal{P}(0)\rangle dt^{\prime},
\label{eq:cf:kappa2}
\end{equation}
where $\delta \mathcal{P}(t) = \mathcal{P}(t)-P$, and $P = \langle \mathcal{P} \rangle$ the ensemble average over time. \\
The viscosity autocorrelation functions are sampled every $5$ fs with a correlation length of $10$ ps. \\

\subsection{Order parameter}
In order to assess the formation and properties of the ice phases, the planar orientational order parameter, $\Psi_k$, can be calculated through\cite{Chaikin1995,Das1999,Crane2008}
\begin{equation}
    \Psi_k = \left|\frac{1}{N}\sum_{j=1}^N \exp^{ik\theta_j}\right|.
\end{equation}
Here, $\theta_j$ is the angle between the vectors of a sample atom and the nearest neighbours. Nearest neighbours are defined as being located at the first maximum of the oxygen-oxygen radial distribution function, $g(r)$. $k$ here takes the value 4, to define a measure of square order, where 1 is high and 0 indicates no order. The order parameter, $\Psi_k$, is evaluated in the $xy$-plane, only.

\section{Results}
\subsection{Diffusivity and shear viscosity}
\label{sec:cf:diff}

\begin{figure}[!h]
	\centering
	\includegraphics[width = 0.8\textwidth]{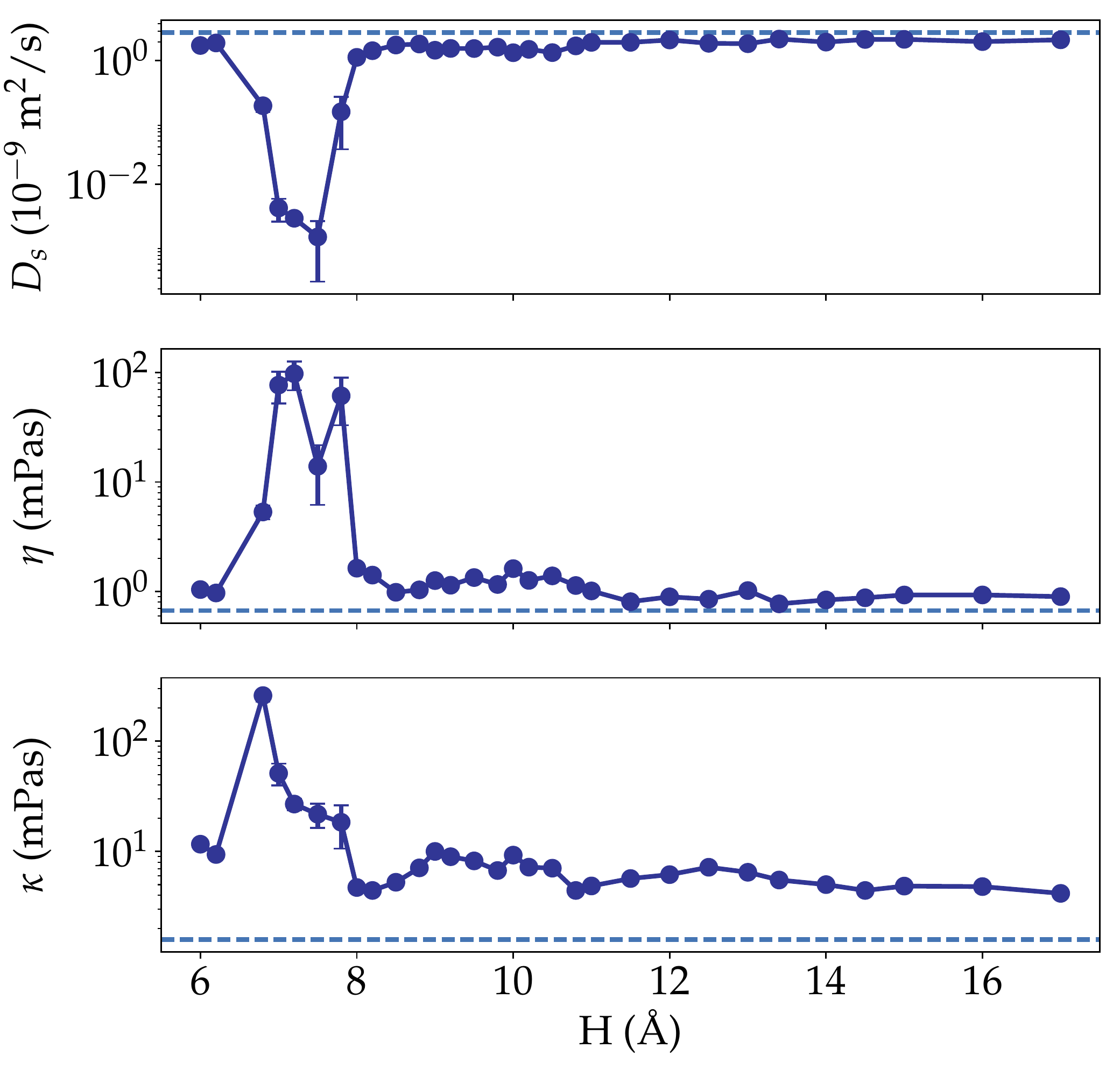}
	\caption{The diffusion coefficient, $D_s$, the shear viscosity, $\eta$, and the bulk viscosity, $\kappa$, as a function of channel separation, $H$. The results are presented on a $y$-log scale. The dashed lines indicate $D_s$, $\eta$ and $\kappa$ for bulk SPC/E water. The water temperature and density are $298$ K and $\rho_{\mathrm{ave}} \simeq 1$ g/cm$^3$, respectively.}
	\label{fig:cf:mob}
\end{figure}

The top panel of Fig. \ref{fig:cf:mob} shows the diffusion coefficient in the $xy$-plane as calculated using the MSD, where $\rho_{\mathrm{ave}} \simeq 1$ g/cm$^3$. Between $6.8 \leq H \leq 7.8$ \AA, the diffusion coefficient is up to $3$ orders of magnitude smaller than in the bulk. This sharp decrease points to a strong hindrance in the mobility. This is further supported by the $2d$ shear viscosity, $\eta$, in the plane parallel to the graphene surface as shown in the middle panel of Fig. \ref{fig:cf:mob} as a function of sheet separation. A sharp increase of the shear viscosity is seen for the same channel heights. Outside this region, whilst $\eta$ at large separations does approach the corresponding value for bulk water at the same density, it nevertheless remains higher at all separations, with some additional enhancement seen for separations of up to $H \simeq 11$ \AA. For the largest separation examined here, $H=40$ \AA, $\eta=0.76 \pm 0.02$ mPas, compared to the bulk value of $\eta=0.67 \pm 0.02$ mPas.\\
When the density is higher, the effect on transport properties is more significant. For instance, for a monolayer of water with $\rho_{\mathrm{ave}} \simeq 1.9$ g/cm$^3$, the diffusion coefficient reaches values as low as $10^{-13}$ m$^2$/s, an order of magnitude lower than for $\rho_{\mathrm{ave}} \simeq 1$ g/cm$^3$ (see Tabs. S2 and S3). Similarly, the shear viscosity is increased by an additional order of magnitude. The shear viscosity shows oscillatory behaviour for channel heights up to 16 \AA. 
\subsection{Bulk viscosity}
The bulk viscosity, $\kappa$, shows similar behaviour to $\eta$, in that there is an increased value observed in confinement (bottom panel of Fig. \ref{fig:cf:mob}). For large separations, $\kappa$ approaches $4$ mPas, which is $2.5$ times larger than the corresponding bulk value, and hence a larger enhancement than recorded for $\eta$. For $6.8 \leq H \leq 7.8$ \AA, whilst an increase compared to larger separations is observed throughout, one separation, $H= 6.8$ \AA, presents as an outlier and shows the largest enhancement by almost a factor of $10$.\\
Overall, the viscosity ratio, $\kappa/\eta$, is larger than in the bulk, with a limiting value of $\kappa/\eta= 6.3 \pm 1.1$ for large separations, compared to $\kappa/\eta=2.3 \pm 0.1$ for bulk water (see Fig. S15). However, for $H=(7-7.8)$ \AA, $\kappa/\eta$ returns to bulk values. These results highlight that the bulk viscosity at moderate confinement is more strongly affected than the shear viscosity, whereas at small separations the properties of the latter are more perturbed. 

\subsection{Ice formation}
\label{sec:cf:struc}
\begin{figure}
	\centering
	\begin{subfigure}[b]{0.4\textwidth}
		\centering
		\includegraphics[width=\textwidth]{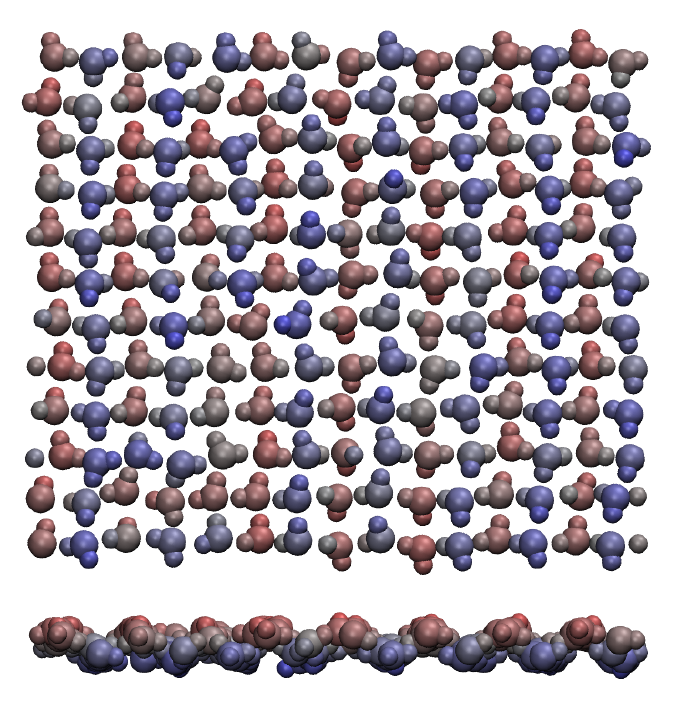}
		\caption{Simulation snapshot}
		\label{fig:cf:ice:sim}
	\end{subfigure}
	\begin{subfigure}[b]{0.5\textwidth}
		\centering
		\includegraphics[width=\textwidth]{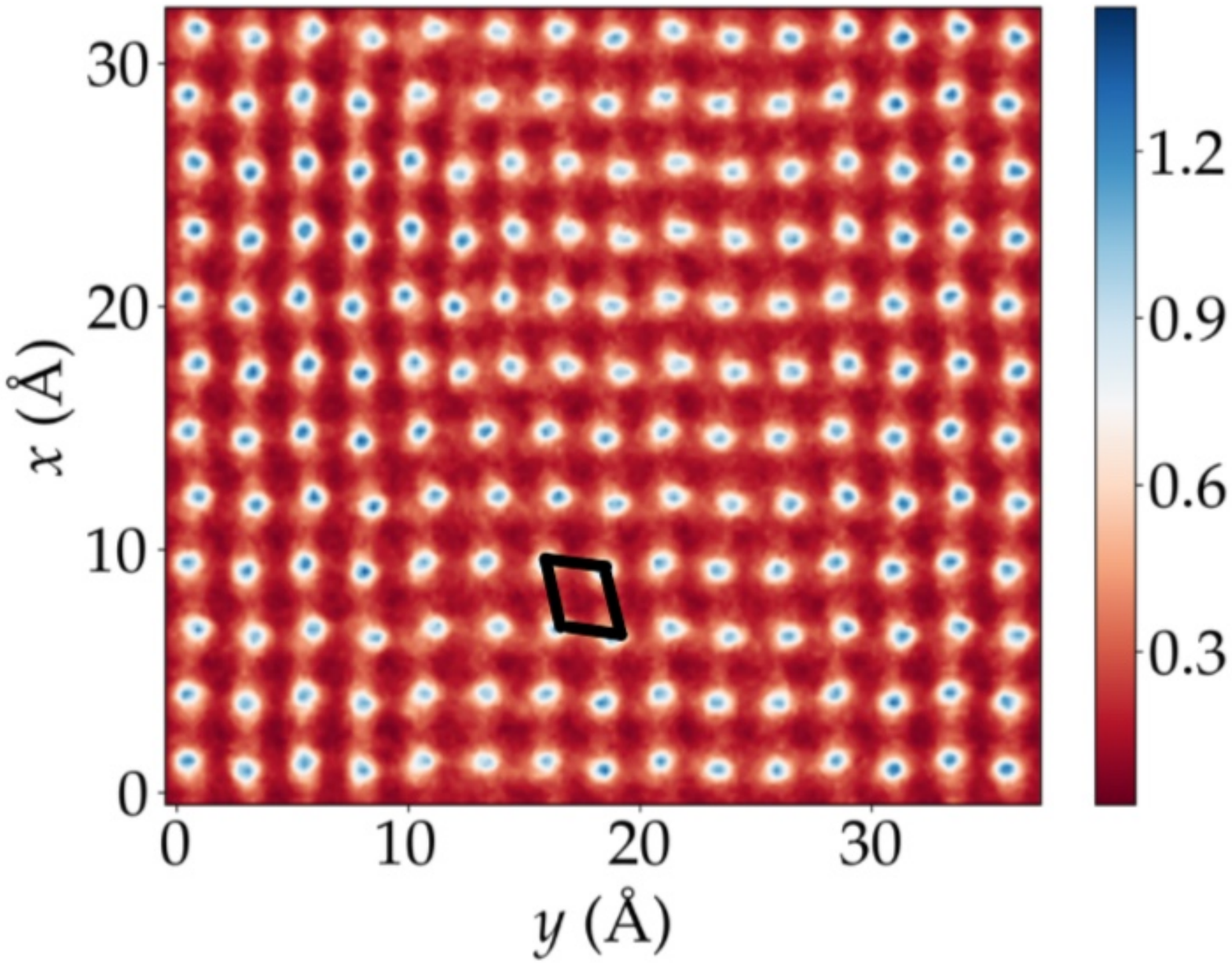}
		\caption{Oxygen density}
		\label{fig:cf:ice:denso}
	\end{subfigure}
	\caption{Sample ice formation for $H=7.5$ \AA~at $T=298$ K and $\rho_{\mathrm{ave}} \simeq 1$ g/cm$^3$. Under this confinement water forms a puckered monolayer ice phase. a) shows a snapshot from MD simulations, both a top and side view, while b) shows the corresponding oxygen density, $\rho_{O} (x,y)$, in the plane. The rhombic lattice is highlighted.}\label{fig:cf:ice}
\end{figure}
\begin{figure}[!h]
	\centering
	\begin{subfigure}[b]{\textwidth}
		\centering
		~~~~\includegraphics[width=0.9\textwidth]{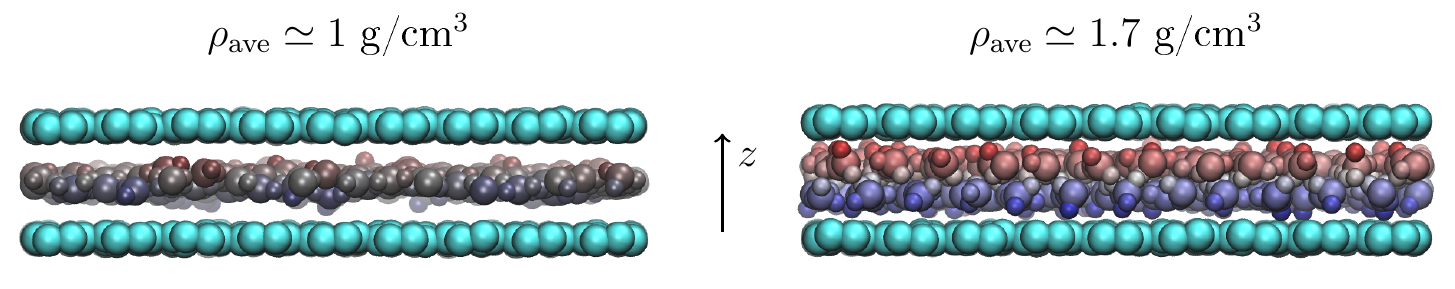}
		\caption{Simulation snapshots}
		\label{fig:cf:rhocomp:md}
	\end{subfigure}
	
	\medskip
	\begin{subfigure}[b]{0.455\textwidth}
		\centering
		\includegraphics[width=\textwidth]{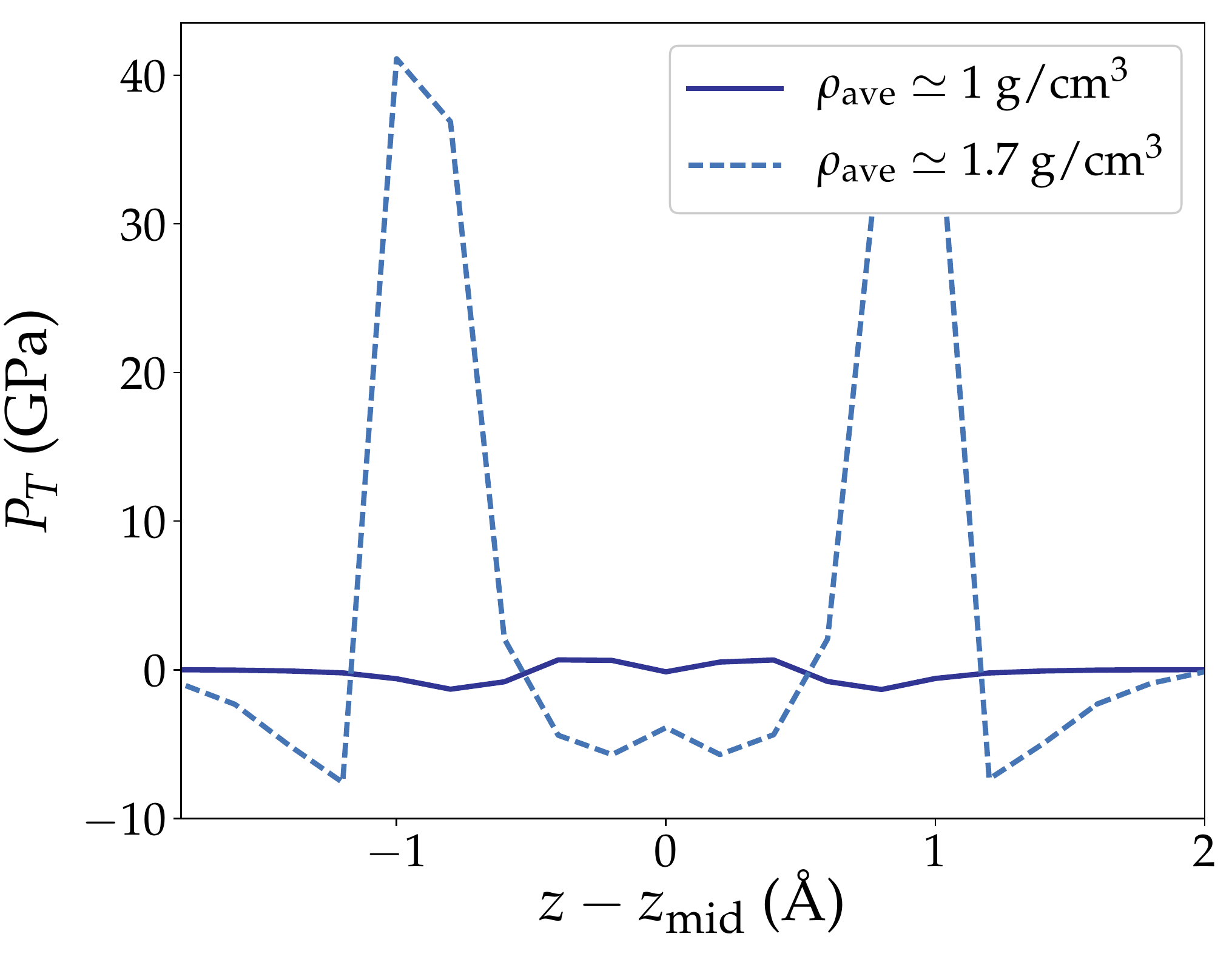}
		\caption{Pressure}
		\label{fig:cf:rhocomp:press}
	\end{subfigure}
	~
	\begin{subfigure}[b]{0.45\textwidth}
		\centering
		\includegraphics[width=\textwidth]{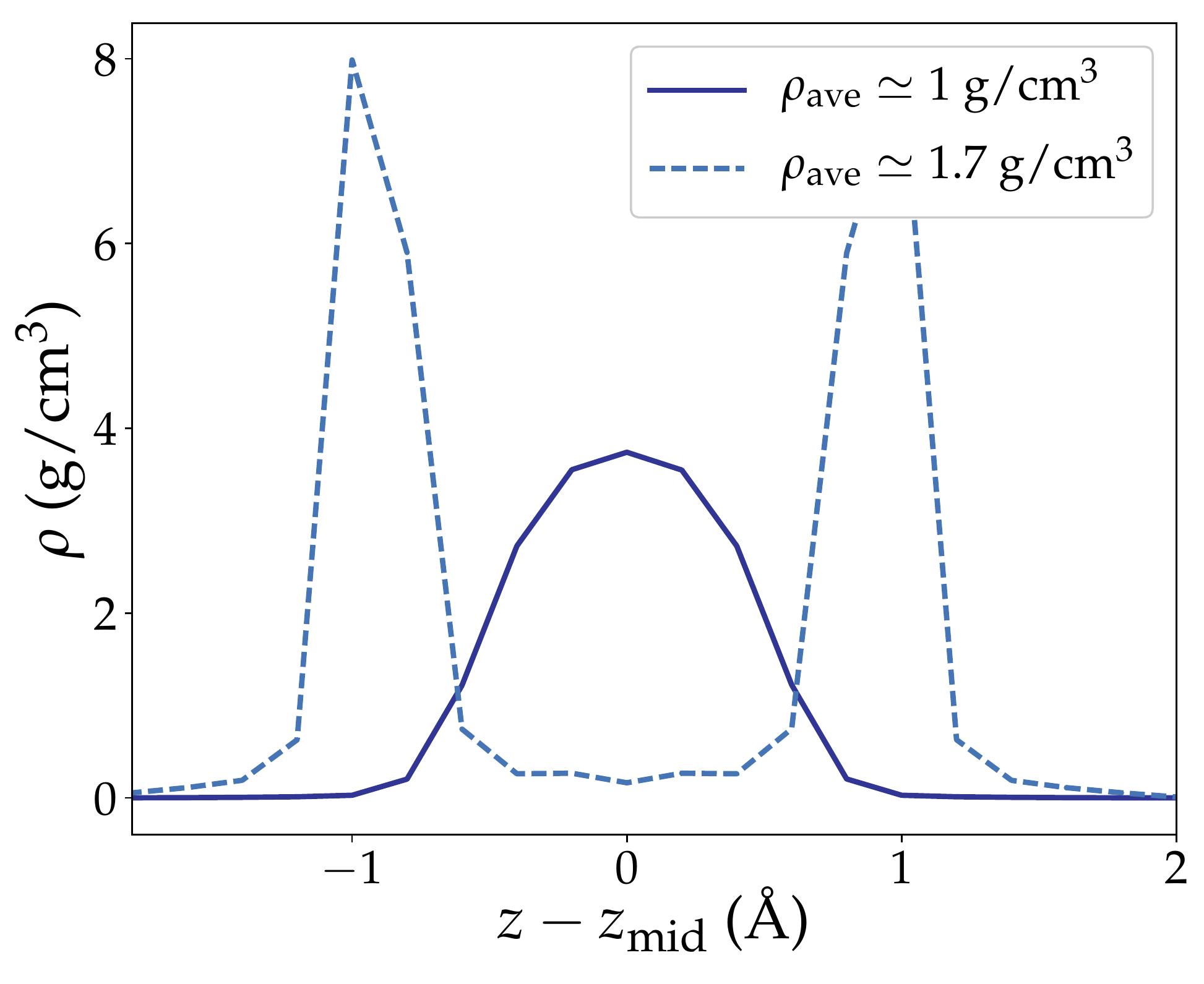}
		\caption{Density}
		\label{fig:cf:rhocomp:rho}
	\end{subfigure}
	\caption{Simulation snapshots (a), tangential pressure (b), $P_{T}(z)$, and density (c), $\rho(z)$, of confined water, where $H=7$ \AA~at $T=298$ K. The average channel densities are $\rho_{\mathrm{ave}} \simeq 1$ and $1.7$ g/cm$^3$, respectively.}\label{fig:cf:rhocomp}
\end{figure}

Our analysis of the simulation snapshots shows that highly ordered monolayer structures form for certain channel heights at room temperature ($298$ K) and bulk densities ($\rho_{\mathrm{ave}} \simeq 1$ g/cm$^3$). For very small distances ($H<6.8$ \AA), a liquid monolayer is observed. However, for $6.8 \leq H \leq 7.8$ \AA, from now on termed the `frozen' regime, ordered systems are found, the nature of which depends on the channel height (Figs. S2-S6).\\
For all separations in the frozen regime, a rhombic structure, also sometimes referred to as square or `square-like' in the literature \cite{Algara-Siller2015,SobrinoFernandezMario2015}, is observed. For smaller distances, the crystalline structure is flat ($H=6.8, 7.0, 7.2$ \AA), whereas for larger separations, a puckered phase is found ($H=7.5, 7.8$ \AA). In Fig. \ref{fig:cf:ice} a sample ice configuration for $H=7.5$ \AA~is shown. As can be seen in Fig. \ref{fig:cf:ice:sim}, the ice phase is predominantly a puckered rhombic monolayer phase. The side view emphasizes the puckered nature of the formation, as highlighted by the colouring. No net polarization is present here, in accordance with previous studies \cite{Mosaddeghi2012,Zhao2014,Zhou2015,SobrinoFernandezMario2015,Yang2017}. Fig. \ref{fig:cf:ice:denso} shows the corresponding $2d$ time-averaged oxygen density, $\rho_{O}(x,y)$, where the rhombic arrangement of the oxygen atoms is evident. \\
Additional studies of this system at higher densities have shown that highly ordered systems form at carbon-to-carbon distances of $6$ \AA~up to distances of $9$ \AA~(Figs. S8-S12). Notably, not only monolayer, but also bilayer ice structures form, with clear bilayer ice formation appearing at widths of $H \geq 7$ \AA. Here, a rhombic ice phase with AB stacking is found. For comparison, the molecular arrangement, density and pressure profiles of the same channel separation, $H=7$ \AA, with densities $\rho_{\mathrm{ave}} \simeq 1$ and $1.7$ g/cm$^3$ are plotted in Fig. \ref{fig:cf:rhocomp} . In the $\rho_{\mathrm{ave}} \simeq 1$ g/cm$^3$ case, the pressures tangential to the graphene sheets, $P_{T}$, are of the order of $1$ GPa, comparable to the pressures in previous studies of ice phases in confinement. The pressures in the higher density system, however, are an order of magnitude larger. The corresponding density profiles are shown in Fig. \ref{fig:cf:rhocomp:rho}. Here, the effect of the increase in the density is clear, with a change from a broadened monolayer to bilayer water. This particular case showcases the sharp morphological changes in the adsorbed solid that occur upon a perturbation in density and highlights the difficulty in mapping the thermodynamic phase to a given geometry. \\

\subsection{Order parameter}
\begin{figure}[!h]
\begin{subfigure}[b]{0.48\textwidth}
	\centering
	\includegraphics[width = \textwidth]{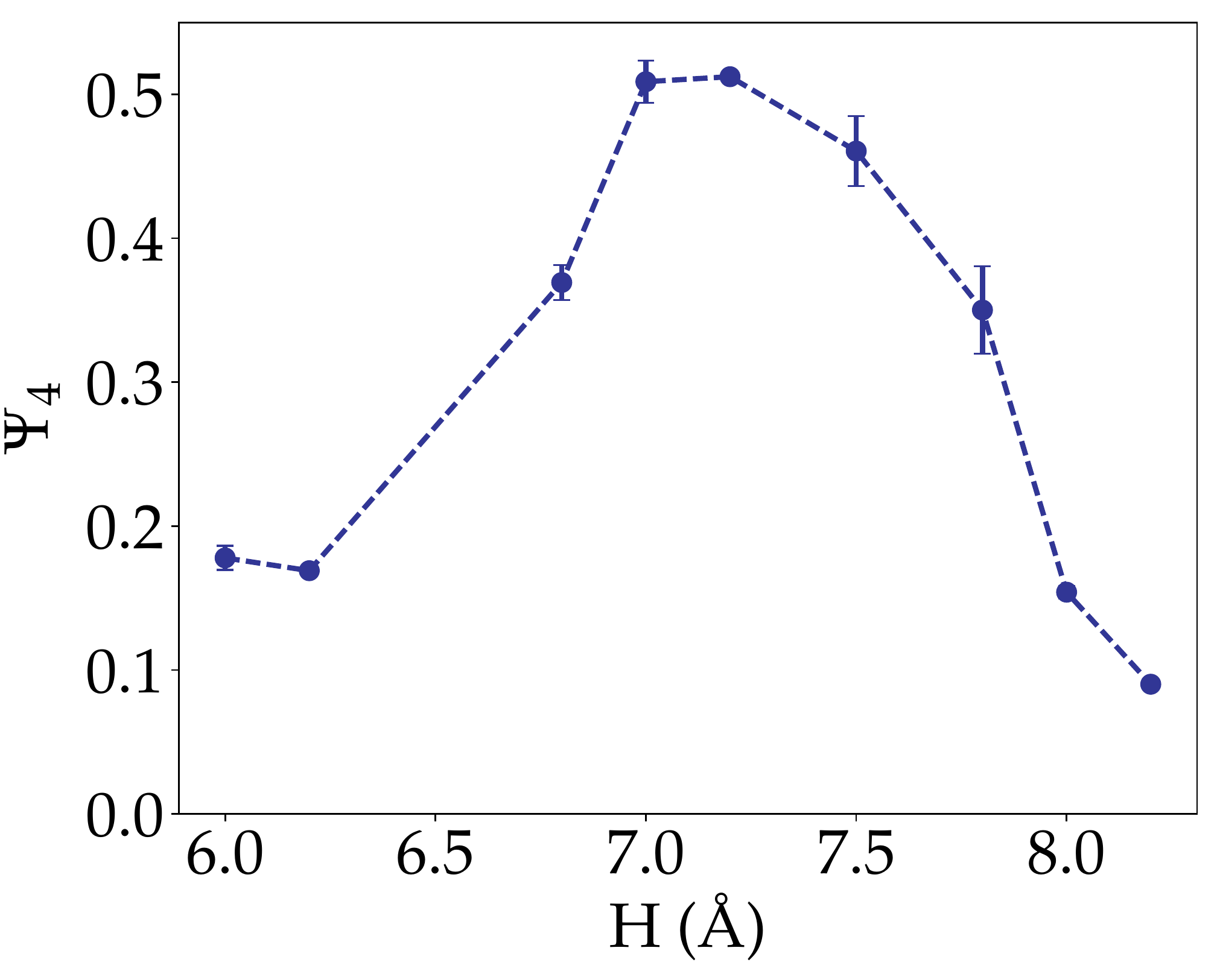}
	\caption{$\rho_{\mathrm{ave}} \simeq 1$ g/cm$^3$}
	\label{fig:cf:op}
\end{subfigure}
~
\begin{subfigure}[b]{0.48\textwidth}
	\centering
	\includegraphics[width = \textwidth]{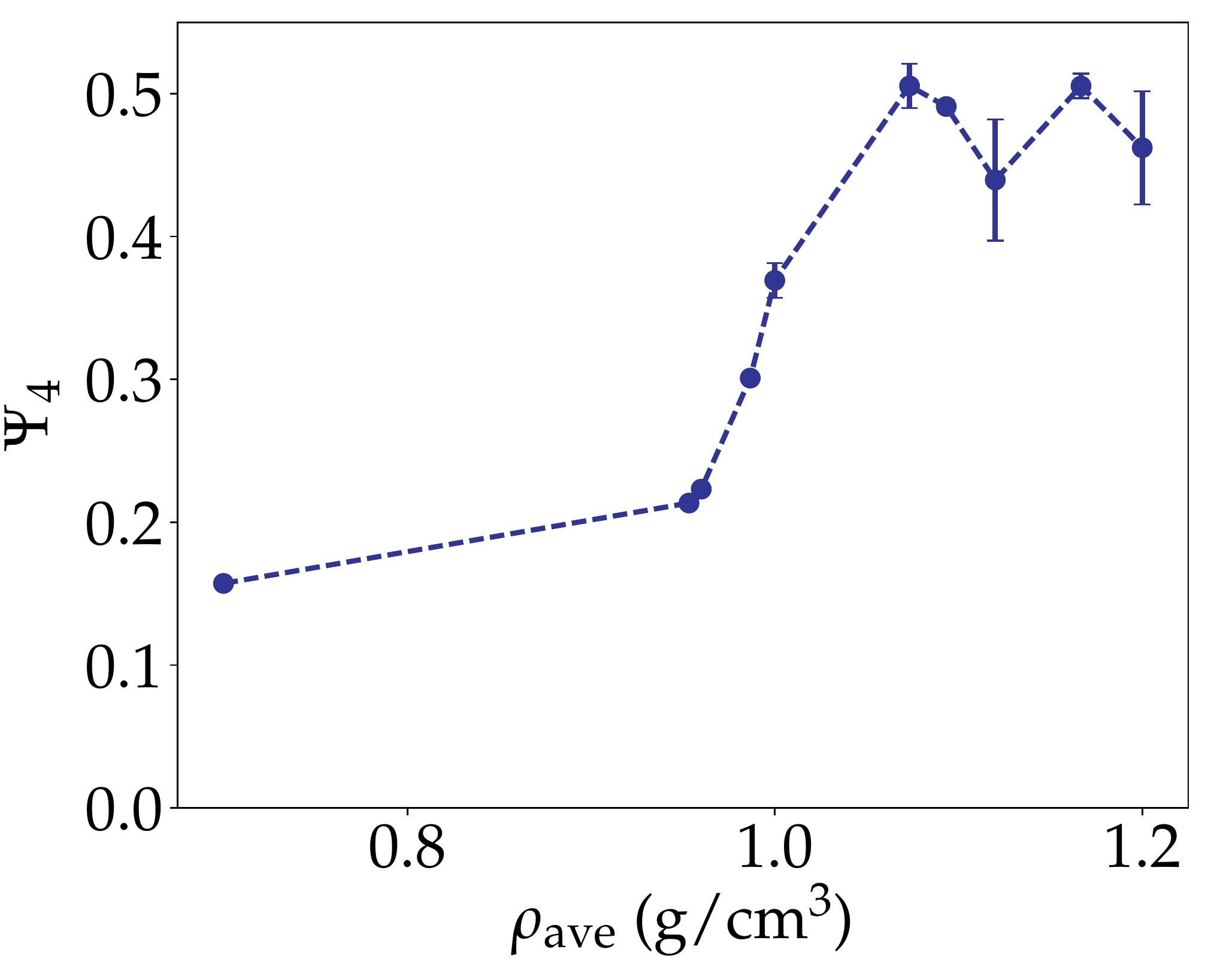}
	\caption{$H=6.8$ \AA}
	\label{fig:cf:op:68}
\end{subfigure}
\caption{Square order parameter, $\Psi_4$, evaluated for water confined between graphene sheets at $T=298$ K. In (a), $\Psi_4$ is calculated at $\rho_{\mathrm{ave}} \simeq 1$ g/cm$^3$ with varying channel heights, with increasing values recorded in the frozen region, $H=(6.8-7.8)$ \AA. In (b), $\Psi_4$ is shown at $H=6.8$ \AA~with varying densities. A phase transition from liquid ($\Psi_4 \sim 0.15$) and ordered ($\Psi_4 \sim 0.5$) is clearly visible by the increase of $\Psi_4$.}
\end{figure}

 Fig. \ref{fig:cf:op} shows the square order parameter, $\Psi_4$, for small channel heights. A clear maximum in the frozen region is seen, indicating increased square order. However, the maximum value achieved is still well below 1, pointing towards a departure from perfect squares. This is in agreement with the appearance of rhombic (rather than square) ice. \\
In order to investigate the unexpected value of $\kappa$ found at $H=6.8$ \AA, the structure and thermodynamic state of water at this confinement is inspected more closely by varying the average density in the channel. Whilst at $\rho_{\mathrm{ave}} \simeq 1$ g/cm$^3$ most of the molecules are part of a rhombic ice structure, a small bubble can also be observed in the monolayer (see Fig. S2). Analysis of the order parameter, $\Psi_4$, at this height (Fig. \ref{fig:cf:op:68}) shows that for densities in the liquid range, $\Psi_4$ is low with values of $\sim$ 0.15, whereas for higher densities, an ordered, square-like system is found ($\Psi_4 \sim 0.5$). The case with an average density of $\rho_{\mathrm{ave}} \simeq 1$ g/cm$^3$ sits in the middle between the two regimes, indicating that the system undergoes a phase transition. At this global density the system is under strain and the normal pressure is negative, with large fluctuations associated with the inherent instability of the phases. \\

\subsection{Radial distribution function}
\begin{figure}[!h]
	\centering
	\includegraphics[width = 0.8\textwidth]{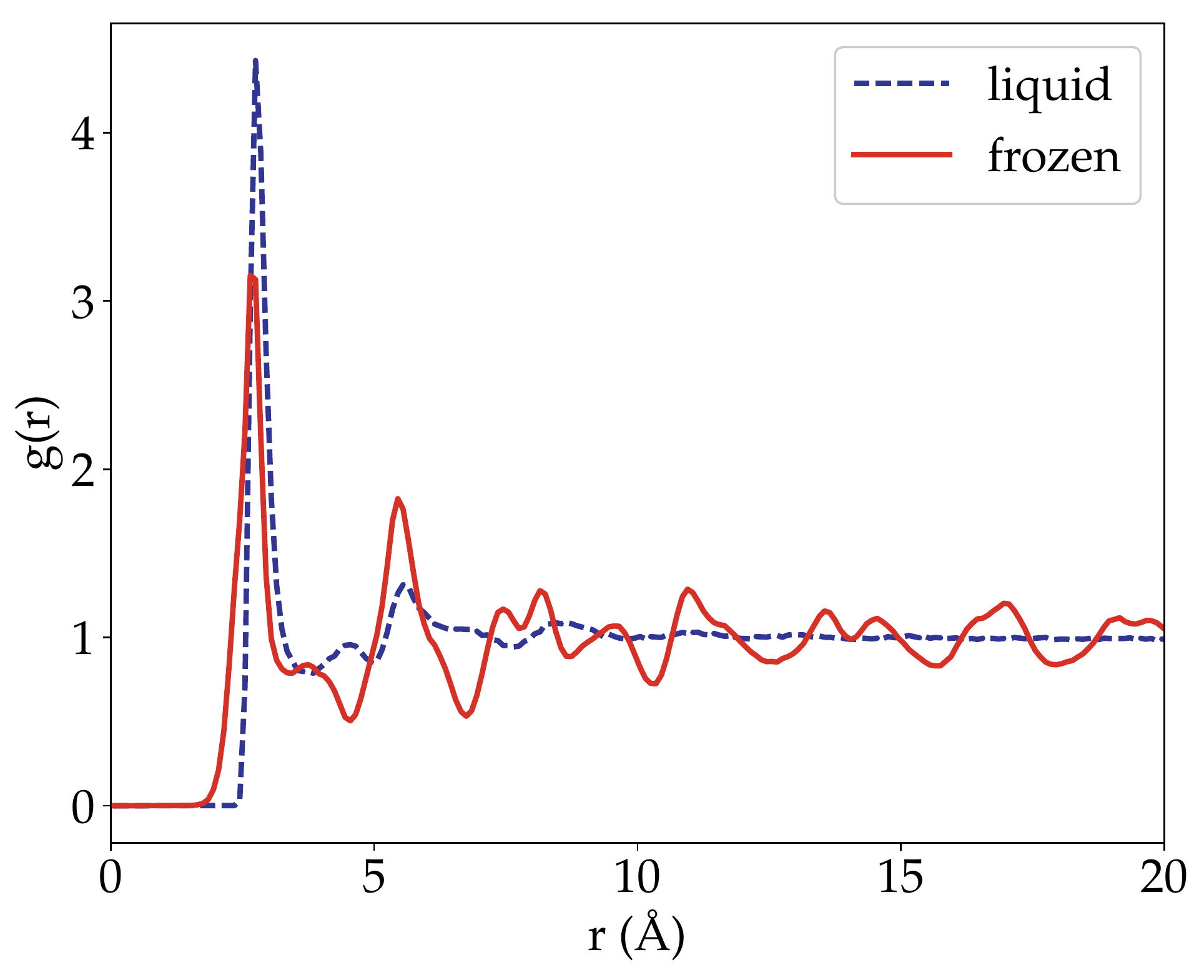}
	\caption{Oxygen-oxygen radial distribution function, $g(r)$, for liquid ($H=6$ \AA, blue, dashed line) and frozen ($H=7.5$ \AA, red, solid line) monolayer water confined between two graphene sheets at $T=298$ K and $\rho_{\mathrm{ave}} \simeq 1$ g/cm$^3$.}
	\label{fig:cf:rdf}
\end{figure}

The two-dimensional oxygen-oxygen radial distribution functions, $g(r)$, in the plane parallel to the graphene sheets ($xy$) for liquid ($H=6$ \AA)  and frozen ($H=7.5$ \AA) monolayer water at $298$ K at an average water density of $1$ g/cm$^3$ are shown in Fig. \ref{fig:cf:rdf}. The liquid monolayer $g(r)$, whilst it shows an additional peak at $\sim 6$ \AA~not present in bulk water, nevertheless converges to $1$ within less than $10$ \AA. The ordered monolayer, on the other hand, shows persistent oscillations over the entire distance examined. The peak of the $g(r)$ is at approximately the same position in both cases, at $r_1 \simeq 2.7$ \AA, an observation that is made for all channel heights, $H$, regardless of ordering. In the crystalline case, a further small peak is found at $r_2 \simeq 3.8$ \AA, with a prominent peak at $r_3 \simeq 5.4$ \AA. $r_2$ is roughly equivalent to $\sqrt{2}r_1$, whilst $r_3 = 2r_1$. This peak spacing is commensurate with a $2d$ square lattice.\\

\subsection{Velocity autocorrelation function}
\begin{figure}[!h]
	\centering
	\includegraphics[width=0.8\textwidth]{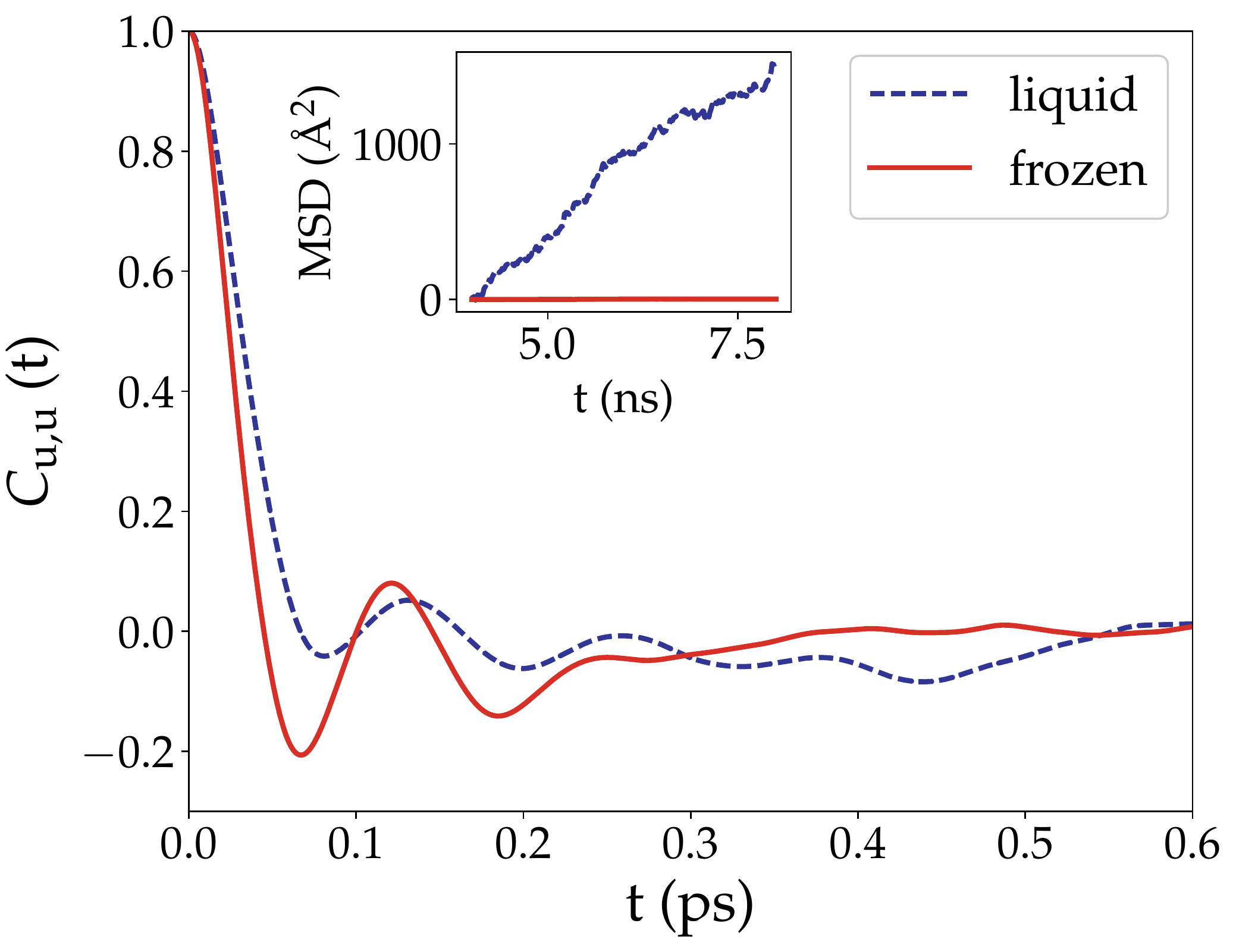}
	\caption{2$d$ velocity autocorrelation functions for confined water at $T=298$ K and $\rho_{\mathrm{ave}} \simeq 1$ g/cm$^3$. The figure shows the difference between a liquid ($H=6$ \AA, blue, dashed line) and frozen ($H=7.5$ \AA, red, solid line) monolayer. The inset shows the mean-squared displacement in the $xy$-plane over the course of the simulation for the liquid and solid phase, respectively.}
	\label{fig:cf:vacf:solid}
\end{figure}

The velocity autocorrelation function (VACF) is a good indicator of long-range order. Fig. \ref{fig:cf:vacf:solid} shows the difference between the 2$d$ VACF for a liquid monolayer ($H=6$ \AA) and an ice monolayer ($H=7.5$ \AA) for room temperature ($298$ K) confined water with average channel densities of $1$ g/cm$^3$. The oscillations in the VACF for the ice phase highlight the ordering, as the change in sign reflects the collisions on short time scales experienced by the water molecules. The VACF for $H=6$ \AA, on the other hand, strongly resembles that of the bulk, though indications of higher ordering prevail here as well, as indicated by a deeper first minimum compared to the bulk. An investigation of the mean-squared displacement (MSD) in the $xy$-plane (Fig. \ref{fig:cf:vacf:solid}, inset) shows results commensurate with this. The MSD of the liquid is of the order of $10^3$ \AA$^2$ during the production run, whereas that of the frozen phase is instead two orders of magnitude smaller and plateaus at long times. This highlights that there is very little motion of the water molecules here associated with the frozen phase over the course of the entire production run, confirming crystallization.\\

\section{Discussion}
The results presented in this work show a significant decrease in mobility of confined water at small channel heights, $6.8 \leq H \leq 7.8$ \AA, when $\rho_{\mathrm{ave}} \simeq 1$ g/cm$^3$, of several orders of magnitude when compared to bulk water. At large separations, $D_s$ and $\eta$ approach the values recorded for bulk SPC/E, though mobility in the plane is always decreased, regardless of the level of confinement. Similar results have been reported in the literature: a comparable study by Neek-Amal et al. \cite{Neek-Amal2016} showed strong oscillations in the shear viscosity, $\eta$, akin to our higher density study, where oscillatory behaviour in the shear viscosity is also observed. At $\rho_{\mathrm{ave}} \simeq 1$ g/cm$^3$, we find a single peak in $\eta$ at small channel heights instead, highlighting the sensitivity of the system to small changes in density and pressure.\\
To the authors' knowledge the bulk viscosity in confinement has only been calculated once in the current literature \cite{Goyal2013}. Notably, water is not studied and no information is given on the phase (fluid or solid) in strong confinement. Similar to the work by Goyal et al., we see an increase in the bulk viscosity when confinement reaches values of a few molecular diameters. A continuous drop in the viscosity ratio, $\kappa/\eta$, is also observed for small separations, indicating that the shear viscosity enhancement is much larger than that of the bulk viscosity in strong confinement. In this work, confinement of less than 3 molecular diameters is additionally studied, which is not covered by Goyal et al. Here, $\kappa$ returns to smaller values and $\kappa/\eta$ in turn increases. \\
In order to further understand the strongly impeded mobility of the water molecules at small channel heights, the phase of water at varying levels of confinement was studied. The results presented demonstrate the formation of ice phases for $6.8 \leq H \leq 7.8$ \AA~when $\rho_{\mathrm{ave}} \simeq 1$ g/cm$^3$, where only a monolayer of water is accommodated between the graphene sheets. Notably, for smaller separations, $H < 6.8$ \AA, a liquid monolayer is observed. In the frozen regime, both flat ($H=6.8, 7.0, 7.2$ \AA) and puckered ($H=7.5,7.8$ \AA) rhombic ice phases with no net polarization are seen, with pressures in the GPa range. \\
These results are qualitatively consistent with previous research on confined water \cite{Zangi2003,Zhao2014,Zhu2015,Yang2017}, where puckered monolayer ice phases are reported for $H\simeq7$ \AA, with flat phases observed for smaller channel heights \cite{Zhu2015,Yang2017}, as reported here. The stability of square and rhombic 2$d$ ice in confinement at GPa pressures has been confirmed using highly accurate diffusion Monte Carlo simulations \cite{Chen2016}. These authors used different methods, including different water force fields and computational approaches, to achieve qualitatively similar results. However, unlike some results presented in the literature \cite{Zhao2014,Yang2017,Zangi2003a,Zhu2015,SobrinoFernandezMario2015}, we do not find bi- or trilayer solid structures at bulk densities. This, however, is consistent with other studies of SPC/E water \cite{Mosaddeghi2012,Algara-Siller2015,Gao2018}, where only at high pressures and densities multilayer ices are reported. \\
The strongly impeded mobility is clearly linked to the phase transition and subsequent solidification of water at small channel heights. Whilst classical molecular dynamics simulations using the SPC/E force field are not as accurate as \textit{ab initio} methods, solidification is nevertheless observed in simulations regardless of force field or methodology \cite{Chen2016}, with differences mostly restricted to the crystal structure that forms. Results for the viscosity ratio further show that crystallization impacts shear and diffusive properties to a larger degree than the bulk viscosity, whilst in the liquid phase the bulk viscosity is more strongly affected by confinement. The investigation of the phase behaviour also allows for an explanation of the large increase in $\kappa$ for $H=6.8$ \AA. As the analysis of the order parameter at different densities shows, water undergoes a liquid-to-solid phase transition at $\rho_{\mathrm{ave}} \simeq 1$ g/cm$^3$, where several phases co-exist. The resulting instability causes the bulk viscosity to increase in this scenario. Similar enhancements are seen for pure, bulk fluids in the vapor-liquid region \cite{Jaeger2018}. Increasing the global density at the same channel height shifts the equilibrium to the solid phase and results in a less pronounced value of $\kappa$ in fully frozen configurations. Whilst the investigations of the impact of $\kappa$ on flow behaviour are beyond the scope of this work, the effects of phase instabilities on transport properties should be an important consideration in future work.\\

\section{Conclusion}
This study bench-marked the dependence of both water structure and transport on the level of confinement experienced by water enclosed between two graphene sheets at $298$ K and average channel densities of $1$ g/cm$^3$. \\
The above discussion demonstrates that transport properties are strongly affected and mobility is impeded in confinement, particularly at nanoscale channel heights. The phase transition and subsequent crystallization of water can account for this phenomenon, where phase changes are accompanied with abrupt changes in the transport properties. Here, the diffusion coefficient and shear viscosity are orders of magnitude smaller and larger, respectively. The phase transition has a particular impact on the bulk viscosity, as highlighted by the strong enhancement found for $H=6.8$ \AA, where the system is unstable.\\
These effects on the properties of water need to be taken into account when trying to optimize the design of nanoscale devices where graphene structures are used as scaffolds, e.g. membranes for water purifications. In particular, the instability of phases needs to be considered in the process and should be either avoided in order to ensure predictable transport behaviour or, on the other hand, could be appropriately exploited. The extreme pressures recorded in confinement can be explained from a molecular point of view \cite{Long2013} and have been suggested as plausible routes for enhancing chemical reactivity in nanopores \cite{Gubbins2018}. Further, regulating the freezing or melting of water requires fine control of the thermodynamic state, namely the density and the pressure.

\begin{acknowledgement}
F.J. was supported through a studentship in the center for Doctoral Training on Theory and Simulation of Materials at Imperial College London funded by the U.K. Engineering and Physical Sciences Research Council (EPSRC) (EP/G036888/1). E.A.M. acknowledges support from EPSRC through research grants to the Molecular Systems Engineering group (Grants EP/E016340, and EP/J014958). Computations were performed employing the resources of the Imperial College High Performance Computing Service and the UK Materials and Molecular Modelling Hub, which is partially funded by EPSRC (EP/P020194/1). 
\end{acknowledgement}

\begin{suppinfo}
Tables of numerical results for the transport properties. \\
Simulation snapshots at different channel heights. \\
Sample input files for molecular dynamics simulations.
\end{suppinfo}

\bibliography{short_lib}

\end{document}